\documentclass[a4paper,twocolumn,10pt]{article}
\usepackage{makeidx}
\usepackage{amsmath}
\usepackage{latexsym}
\usepackage{amssymb}

\usepackage{graphicx}
\usepackage{float}

\begin{document}
\twocolumn[%
\newlength{\abswidth}
\setlength{\abswidth}{\linewidth}
\addtolength{\abswidth}{-2in}
\hbox to 16.5 true cm{\hfill TOKAI-HEP/TH-04-09-08\ \ \ \ \ \ \ }
\begin{center}
{\large\bf Dilatonic Inflation and SUSY Breaking in Modular-Invariant Supergravity${}^*$}
\\
\vspace{1.5\baselineskip}
Mitsuo J. Hayashi${}^\dag$ and Tomoki Watanabe
\\
\vspace{\baselineskip}
{\it Department of Physics, Tokai University, 1117 Kitakaname, Hiratsuka, 259-1292, Japan}
\\
\vspace{4pt}
\footnotesize{E-mail: ${}^\dag$mhayashi@keyaki.cc.u-tokai.ac.jp\\
${}^*$Talk at ICHEP2004, BICC at Beijing, Aug. 16-22, 2004}
\vspace{\baselineskip}

{\bf{Abstract}\\\vspace{\baselineskip}}
\begin{minipage}{\abswidth}
The theory of inflation will be investigated as well as supersymmetry breaking in the context of supergravity, incorporating the target-space duality and the non-perturbative gaugino condensation in the hidden sector. 
The inflation and supersymmetry breaking occur at once by the interplay between the dilaton field as inflaton and the condensate gauge-singlet field. The model satisfies the slow-roll condition which solves the $\eta$-problem. When the particle rolls down along the minimized trajectry of the potential $V(S,Y)$ at a duality invariant fixed point $T=1$, we can obtain the $e$-fold value $\sim 57$. And then the cosmological parameters obtained from our model well match with the recent WMAP data combined with other experiments. 
This observation is suggesting to consider the string-inspired supergravity as the fundamental of the theory of the evolution of the universe as well as the particle theory.
\end{minipage}
\vspace{\baselineskip}
\end{center}
]

{\large\bf 1. Inflationary Cosmology\vspace{12pt}}

On February, 2003, WMAP combined with the Other Experiments had shown that the big bang and inflation theories continue to be true\cite{ref:1}.
Inflation models introduce scalar field(s) called inflaton, which must satisfy:  the slow-roll condition that should predict the Number of e-folds, spectral index and its running as well as the the spectrum of the density perturbation of the CMB anisotropy.

There are several problems in constructing theories of inflationary universe: 1) What is it, the inflaton? 2) What kind of theoretical frame works is the most appropriate as the theory of particle physics, inflation and recently observed accelerating universe? 3) How to explain the contents of the universe?: 
Baryonic Matter 4\%, \ \ Dark Matter 23\%, \ \ Dark Energy 73\%, and so on.

It seemed require far richer structures of contents than those of standard Theory of Particles.

Here we concentrate on a supergravity inspired by superstrings. The well-known difficulty of supergravity is that the potential form gave the $\eta$-problem, which breaks the slow-roll condition. The string-inspired supergravity which was derived from the $d=10$ heterotic string dimensionally reducted to
N=1, d = 4 supergravity, whose typical features are: i) No-scale structure at the tree level. ii) $E_8\times E_8$ gauge group (one of the $E_8$ is called the hidden sector of the gauge group). iii) Non-perturbative gaugino condensation in the hidden sector can break the supersymmetry. iv) Modular invariance, acting on a single modulus  $T$, valid at any string-loop order (Target-space duality).    

We would like to show that the inflation and the supersymmetry breaking occur at once by the interplay between the dilaton field as inflaton and the condensate gauge-singlet field rolling down the inflationary trajectory, free from the $\eta$-problem\cite{ref:2}.

\vspace{12pt}
{\large\bf 2. A String-inspired Supergravity\vspace{12pt}}

The most general form of Lagrangian in $N=1$ Supergravity at the tree-level is\cite{ref:3}:
\begin{align*}
\mathcal{L}&=-\frac{1}{2}\left[e^{-K/3}S_0\bar{S}_0\right]_D \\
&+\left[S_0^3W\right]_F +\left[f_{ab}W^a_\alpha \epsilon^{\alpha\beta}W^b_\beta\right]_F,
\end{align*}
where the K\"{a}hler potential $K$ is given by
$$
K=-\ln \left(S+S^\ast\right)-3\ln \left(T+T^\ast-|\Phi_i|^2\right),
$$
and the gauge function $f_{ab}$ is
$$
f_{ab}=\delta_{ab} S.
$$
In order to construct the effective theory of gaugino condensation, we introduce the composite superfield $U$ of the gaugino condensation\cite{ref:4}:
$$
U=\delta_{ab}W^a_\alpha \epsilon^{\alpha\beta}W^b_\beta/S_0^3
=(\lambda\lambda+\cdots)/S_0^3
$$
where $\lambda$ is the gaugino fields in the Hidden sector.
We may construct the effective theory of gaugino condensation. As $U$ has conformal weight 3, we will introduce the superfield:
$$
\tilde{U}=U/S_0^3,
$$
of conformal weight 0. Then, in terms of $\tilde{U}$, the gauge kinetic term of eq.(1) gives the contribution to the tree-level superpotential:
$$
\tilde{W}^{\rm tree}=S\tilde{U}.
$$

The one-loop modification of the superpotential should be $S$ independent and, of the form:
$$
\tilde{W}^{\rm 1-loop}=\tilde{U}f(\tilde{U},T),
$$
with $f$ a modular function of weight 0, while $\tilde{U}$ has a modular form of weight $-3$. Following the general arguments based on the fulfilment of anomalous Ward identities, $f$ is required to contain the term:
$$
\frac{\beta_0}{96\pi^2}\log{\tilde{U}},
$$
where $\beta_0$ is the coefficient of $\beta$-function.
The effective gauge kinetic function is given by:
$$
f=S+3b\ln[Y\eta^2(T)]+Const.
$$

The effective K\"{a}hler potential and superpotential incorporating modular invariant 1-loop corrections are given as\cite{ref:4}:
\begin{eqnarray}
K&=&-\ln\left(S+S^\ast\right)
\nonumber\\
&&{}-3\ln\left(T+T^\ast-|Y|^2-|\Phi_i|^2\right)
\end{eqnarray}
and
\begin{equation}
W=3bY^3\ln\left[c\>e^{S/3b}\>Y\eta^2(T)\right]+W_{\rm matter}
\end{equation}
where $\eta$ is the Dedekind's $\eta$ function, $c$ is a free parameter in the theory and $b=\frac{\beta_0}{96\pi^2}$ ($\beta_0$ is the 1-loop beta-function coeffiients).

Since $\langle S+S^\ast\rangle =\alpha^\prime m_{\rm pl}^2$, the choice:
\begin{equation}
[e^{-K/3}S_0\bar{S}_0]_{\theta=\bar\theta=0}=[S+\bar{S}]_{\theta=\bar\theta=0}
\end{equation}
is corresponded to the conventional normalization of the gravitational action:
\begin{equation}
\mathcal{L}_{\rm grav} \sim [e^{-K/3}S_0\bar{S}_0]_{\theta=\bar\theta=0}R.
\end{equation}

Then, the scalar potential is as follows:
\begin{eqnarray}
&&V(S,T,Y)=
\frac{3(S+S^\ast)|Y|^4}{(T+T^\ast-|Y|^2)^2}
\nonumber\\ 
&&\Bigg(
3b^2 \left|1+3\ln\left[c\>e^{S/3b}\>Y\eta^2(T)\right] \right|^2
\nonumber\\ 
&&+\frac{|Y|^2}{T+T^\ast-|Y|^2}
\Bigg|S+S^\ast
\nonumber\\
&&-3b\ln\left[c\>e^{S/3b}\>Y\eta^2(T)\right] \Bigg|^2
\nonumber\\ 
&&+6b^2|Y|^2
\Bigg[2(T+T^\ast)
\left|\frac{\eta^\prime (T)}{\eta(T)}\right|^2
\nonumber\\
&&+\frac{\eta^\prime (T)}{\eta(T)}
 +\left(\frac{\eta^\prime (T)}{\eta(T)}\right)^\ast \Bigg]
 \Bigg),
\end{eqnarray}
where matter fields are neglected and $Y$ is defined by
$
\tilde{U}=U/S_0^3=Y^3.
$

\vspace{12pt}
{\large\bf 3. Inflationary Trajectory\vspace{12pt}}

The potential is modular invariant and shown to be stationary at the self-dual points $T=1$ and $T=e^{i\pi /6}$. 

\begin{figure}[H]
\begin{center}
\includegraphics[scale=0.8]{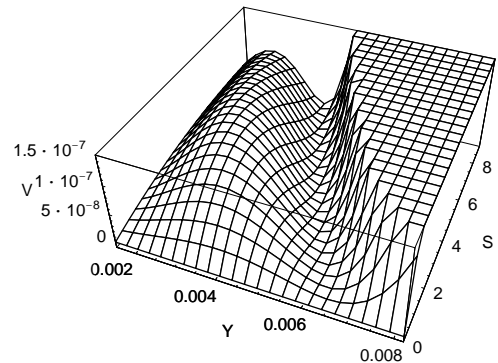}
\end{center}
\caption{The plot of $V(S,Y)$ at fixed $T=1$ (self-dual point) with $c=183,\ b=5.5$.}
\end{figure}
We found that the potential $V(S,Y)$ at $T=1$ has a stable minimum. (See Fig.1).
The stable minimum of $V_Y(S)=0$ and a saddle point exist.
We can see a valley of the potential and a stable minimum of  
$V_Y(S)=0$ at $(Y_{min},S_{min})\sim (0.00646, 0.435)$. 

Therefore, we may conclude that
inflation arises
by the evolution of dilaton field $S$
and supersymmetry is broken
by the condensated field $Y$,
 provided it begins at the unstable saddle point and slowly rolls down to the minimum.

The inflationary trajectory will be well approximated by the equation:
\begin{equation}
Y_{\rm min}(S)\sim 0.00663e^{-S/16.2}.
\end{equation}

\begin{figure}[H]
\begin{center}
\includegraphics[scale=0.8]{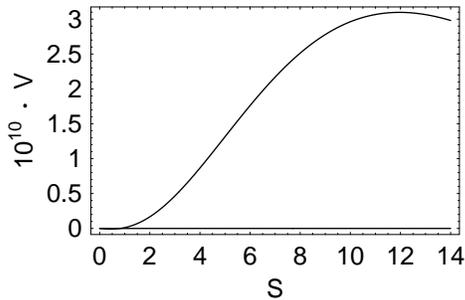}
\end{center}
\caption{The plot of $V(S)$ minimized with respect to $Y$. The minimum value of the potential is $V(S_{\rm min})\sim -9.3\times 10^{-13}$.}
\end{figure}

In Fig. 2, we have shown a plot of $V(S)$ minimized with respect to $Y$.
As shown by Ferrara {\it et al.}, supersymmmetry is broken by the hidden sector gaugino condensation because 
$\langle |F| \rangle \propto \langle |\lambda\lambda |\rangle\neq 0$.

Our main purpose of this paper is to prove that the dilaton field plays the role of inflaton field.

The slow-roll parameters (in Planck units $m_{\rm Pl}/\sqrt{8\pi}=1$) are defined by:
\begin{equation}
\epsilon_\alpha=\frac{1}{2}\left(\frac{\partial_\alpha V}{V}\right)^2
\quad ,\quad 
\eta_{\alpha\beta}=\frac{\partial_\alpha\partial_\beta V}{V}.
\end{equation}
The slow-roll condition demands both values to be lower than 1. It is the end of inflation, when the slow-roll parameter $\epsilon_\alpha$ reaches the value 1. After passing through the end of inflation, ``matters" may be produced during the oscillations around the minimum of the potential (reheating) with the critical density, i.e. $\Omega=1$.

\begin{figure}[H]
\begin{center}
\includegraphics[scale=0.8]{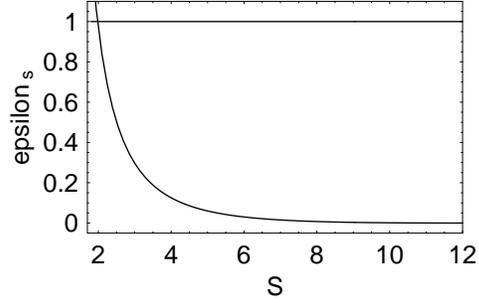}
\caption{The plot of $\epsilon_S$.}
\end{center}
\end{figure}

\begin{figure}[H]
\begin{center}
\includegraphics[scale=0.8]{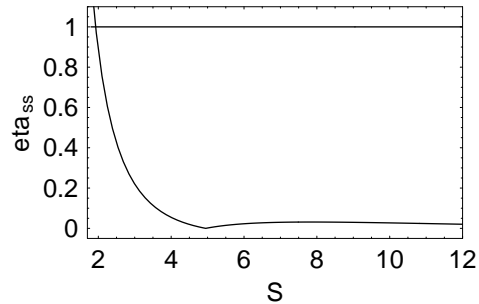}
\caption{The plot of $|\eta_{SS}|$.}
\end{center}
\end{figure}

The values of $\epsilon_S$ and $\eta_{SS}$ are obtained numerically in Figs. 3 and 4 fixing the parameters $c=183\ {\rm and}\ b=5.5$; we find the condition is well satisfied, and the $\eta$-problem can just be avoided.

Number of $e$-folds at which a comoving scale $k$ crosses the Hubble scale $aH$ during inflation is given by:
\begin{align*}
N(k)&\sim 62-\ln\frac{k}{a_0H_0} 
-\frac{1}{4}\ln\frac{(10^{16}\>{\rm GeV})^4}{V_k} \\
& +\frac{1}{4}\ln\frac{V_k}{V_{\rm end}},
\end{align*}
where we assume $V_{\rm end}=\rho_{\rm reh}$. We focus the scale $k_*=0.05\> {\rm Mpc^{-1}}$ and the inflationary energy scale is 
$V\sim10^{-10}\sim(10^{16}\>{\rm GeV})^4$ 
as shown in Fig. 2, therefore the number of $e$-folds which corresponds to our scale must be around 57. 

On the other hand, using the slow-roll approximation, $N$ is also calculated by:
\begin{equation}
N\sim -\int^{S_2}_{S_1}\frac{V}{\partial V}dS.
\end{equation}
We could have obtained the number $\sim 57$, by integrating from  
$S_{\rm end}\sim 1.98$ to $S_*\sim 10.46$,
 fixing the parameters $c=183\ {\rm and}\ b=5.5$,  
 i.e. our potential has the ability to produce the cosmologically plausible number of $e$-folds. Here $S_*$ is the value corresponding to $k_*$.

Next, a scalar spectral index standing for a scale dependence of the spectrum of density perturbation and its running are defined by:
\begin{align*}
n_s-1&=\frac{d\ln \mathcal{P_R}}{d\ln k} \\
\alpha_s&=\frac{dn_s}{d\ln k}.
\end{align*}
These are approximated in the slow-roll paradigm as:
\begin{align*}
n_s(S)&\sim 1-6\epsilon_S+2\eta_{SS} \\
\alpha_s(S)&\sim 16\epsilon_S\eta_{SS} -24\epsilon_S^2-2\xi^2_{(3)},
\end{align*}
where $\xi_{(3)}$ is an extra slow-roll parameter that includes trivial third derivative of the potential.  
Substituting $S_*$ into these, we have $n_{s*}\sim0.95$ and $\alpha_{s*}\sim -4\times10^{-4}$. 

Because $n_s$ is not equal to 1 and $\alpha_{s}$ is negligible, our model supports the model with tilted power law spectrum. The value of $n_{s*}$ is consistent with the recent observations; 
 the best fitting of them (WMAPext, 2dFGRS and Lyman $\alpha$) for power law $\Lambda$CDM model suggests
  $n_s(k_*)=0.96\pm0.02$.

Finally, estimating the spectrum of the density perturbation caused by slow-rolling dilaton:
\begin{equation}
\mathcal{P_R}\sim\frac{1}{12\pi^2}\frac{V^3}{\partial V^2},
\end{equation}
we find $\mathcal{P_R}_*\sim2.1\times10^{-9}$. 

This result matches the measurements as well. Incidentally speaking, the energy scale $V\sim10^{-10}$ is also within the constrained range by Liddle and Leach\cite{ref:5}. 

\vspace{12pt}
{\large\bf 4. Conclusion and Summary\vspace{12pt}}

 Now we conclude that inflation and supersymmetry breaking occur at once by the interplay between the dilaton field as inflaton and the condensate gauge-singlet field. Since the inflaton field is concerned with the Planck scale physics, the dilaton field seems the most presumable candidate of the inflaton. 
 Among the possibe supergravity models of inflation, the modular invariant model here revisited seems open the hope to construct the realistic theory of particles and cosmology, including the undetected objects, such as the inflaton, the dark matter and dark energy.
 
 Finally we will mention on the alternative possibility to explain the accellated expansion of the universe in the framework of Poincar\'e gauge theory of genaral relativity. For the Dirac and the Rarita-Schwinger fields, the correction to the energy momentum tensor due to spin can provide the negative pressure and be an alternative to false vacuum in the early stage of the universe \cite{ref:6}.

\end{document}